\documentclass[conference,compsoc]{IEEEtran}
\usepackage{cite}
\ifCLASSINFOpdf
\usepackage[pdftex]{graphicx}
\else
\fi

\def\aj{\emph{AJ.}}

\def\apj{\emph{ApJ.}}

\def\apjl{\emph{ApJ. Lett.}}
\def\araa{\emph{ARA\&A}}
\def\mnras{\emph{MNRAS}}
\def\nat{\emph{Nature}}

\IEEEoverridecommandlockouts\IEEEpubid{\makebox[\columnwidth]{U.S. Government work not protected by U.S. copyright \hfill} \hspace{\columnsep}\makebox[\columnwidth]{ }}

\begin{document}
\title{Integrating Temporal and Spectral Features of Astronomical Data Using Wavelet Analysis for Source Classification}

\author{\IEEEauthorblockN{Tilan N. Ukwatta}
\IEEEauthorblockA{Director's Postdoctoral Fellow,\\ Space and Remote Sensing (ISR-2)\\
Los Alamos National Laboratory,\\ Los Alamos, NM 87545, USA.\\
Email: tilan@lanl.gov}
\and
\IEEEauthorblockN{Przemek R. Wozniak}
\IEEEauthorblockA{Space and Remote Sensing (ISR-2)\\
Los Alamos National Laboratory,\\ Los Alamos, NM 87545, USA.\\
Email: wozniak@lanl.gov}
}

%Telephone: (800) 555--1212, Fax: (888) 555--1212}

%\IEEEspecialpapernotice{(U.S. Government work not protected by U.S. copyright)}

%\IEEEpubid{U.S. Government work not protected by U.S. copyright}

\maketitle

\begin{abstract}
Temporal and spectral information extracted from a stream of photons received
from astronomical sources is the foundation on which we build understanding
of various objects and processes in the Universe.
Typically astronomers fit a number of models separately to light curves and
spectra to extract relevant features. These features are then used to classify, identify,
and understand the nature of the sources. However, these feature extraction methods
may not be optimally sensitive to unknown properties of light curves and
spectra. One can use the raw light curves and spectra as features to train
classifiers, but this typically increases the dimensionality of the problem,
often by several orders of magnitude. We overcome this problem by integrating
light curves and spectra to create an abstract image and using wavelet analysis
to extract important features from the image. Such features incorporate both
temporal and spectral properties of the astronomical data. Classification is
then performed on those abstract features. In order to demonstrate this
technique, we have used gamma-ray burst (GRB) data from the NASA's Swift
mission to classify GRBs into high- and low-redshift groups. Reliable selection
of high-redshift GRBs is of considerable interest in astrophysics and cosmology.
\end{abstract}


% \fi
\IEEEpeerreviewmaketitle

\section{Introduction} \label{introduction}
In astronomy and astrophysics,
information extracted from a stream of photons is used to
understand various objects and processes. Generally
three types of information are extracted from these photons:
their direction, arrival time and energy. Using these observables
astronomers construct sky maps, light curves and spectra.
We try to understand the underlying astrophysical processes
and their sources by fitting various theoretical or empirical
models to these light curves and spectra separately.
These fits allow us to extract relevant features used to characterize sources.
Before model fitting, the light curves and spectra may be integrated over
selected energy bands and time intervals.
Such integration may remove potentially valuable
information on the time and energy evolution of the phenomenon.

Classification is usually one of the first tasks performed to gain
insight into a previously unknown phenomenon. Historically,
this had been accomplished by looking at a handful of ad-hoc features,
possibly extracted from light curve and spectral models,
and trying to identify clusters or groups in low-dimensional projections
of the data. This is a relatively straightforward process if there
is an obvious clustering. However, in most cases the clustering is only evident
in multi-dimensional parameter spaces and requires a more rigorous machine-learning
approach. Indeed many attempts have been
made to perform such classification~\cite{Morgan2012,Sharma2014,Wright2015, Miller2015}.
For these machine-learned classifications, features were
extracted by fitting energy integrated light curves and
time integrated spectra. To avoid loosing valuable information,
one can consider using raw light curves and spectra as features to train
classifiers. But this increases the dimensionality of the problem by many
orders of magnitude and typically reduces the performance of classifiers.

Here we introduce a novel method that can be used to
potentially harness most of the information present
in photon streams from astrophysical sources. As a case study,
we apply our methodology for tackling a challenging
task of classifying high-energy astrophysical transient
phenomenon known as gamma-ray bursts (GRBs) into a high-redshift and a low-redshift
classes only based on promptly available high energy data, as opposed
to ``expensive'' low energy follow-up measurements that take time to collect.
In section~\ref{grb} we introduce the GRB phenomenon and the classification
problem that we are going to address.
We describe our methodology using GRB data in section~\ref{methodology}.
Our results are presented in section~\ref{results}. In section~\ref{discussion}
we discuss our results and applicability of the method to other
astrophysical classification problems. Finally, we give our conclusions in
section~\ref{conclusions}.

\section{Gamma-ray Bursts} \label{grb}
Gamma-ray bursts are often characterized as the most energetic electromagnetic
explosions since the beginning of the Universe.
They are normally first detected in prompt gamma-ray emission phase
followed by the afterglow emission in X-ray, optical and in some cases radio
energy bands~\cite{Gehrels2008,Gehrels2009}.
The prompt gamma-ray emission from GRBs shows very complicated
time profiles. Based on the duration and spectrum, two classes of
bursts have been observed: those that last less
than two seconds and have on average hard spectra (short GRBs),
and those that last longer than two seconds and are spectrally
softer (long GRBs). The exact nature of the GRB progenitors is
unknown, although it is possible that long GRBs come from the
collapse of massive, rapidly rotating stars~\cite{Woosley2006a,Woosley2006b}
and short GRBs result from the merger of compact objects such as neutron
stars and black holes~\cite{Eichler1989,Narayan1992}.
Regardless of the progenitor system, accretion onto the
resulting compact object is thought to create a highly relativistic jet.
The prompt gamma-ray emission from the GRBs arises from the internal shocks due
to collisions within the jet between faster shells of material ejected at later times
with slower shells ejected earlier ~\cite{Rees1994}. The afterglow emission
is generated when the shock wave collides and interacts with the interstellar medium~\cite{Gehrels2008}.

As record holders for the apparent brightness of the electromagnetic emission,
GRBs and their afterglows should be detectable out to
redshift of z $>$ 10~\cite{Lamb2000}. Therefore, the study of
high-redshift GRBs (hereafter high-z GRBs) offers a technique to probe
the early Universe during the epoch of re-ionization, the early star formation
and evolution, and the metal enrichment history of the Universe~\cite{Lamb2000}.

The $Swift$ Gamma-Ray Burst Mission~\cite{Gehrels2004} opened a new window
to explore the high redshift Universe using GRBs.  The highest
spectroscopically confirmed redshift GRB detected thus far is GRB 090423 with
z=8.2~\cite{Tanvir2009,Salvaterra2009}.
The highest photometrically measured burst is GRB 090429B with a redshift
of 9.4~\cite{Cucchiara2011}.

\begin{figure}[!t]
\centering
\includegraphics[width=3.5in]{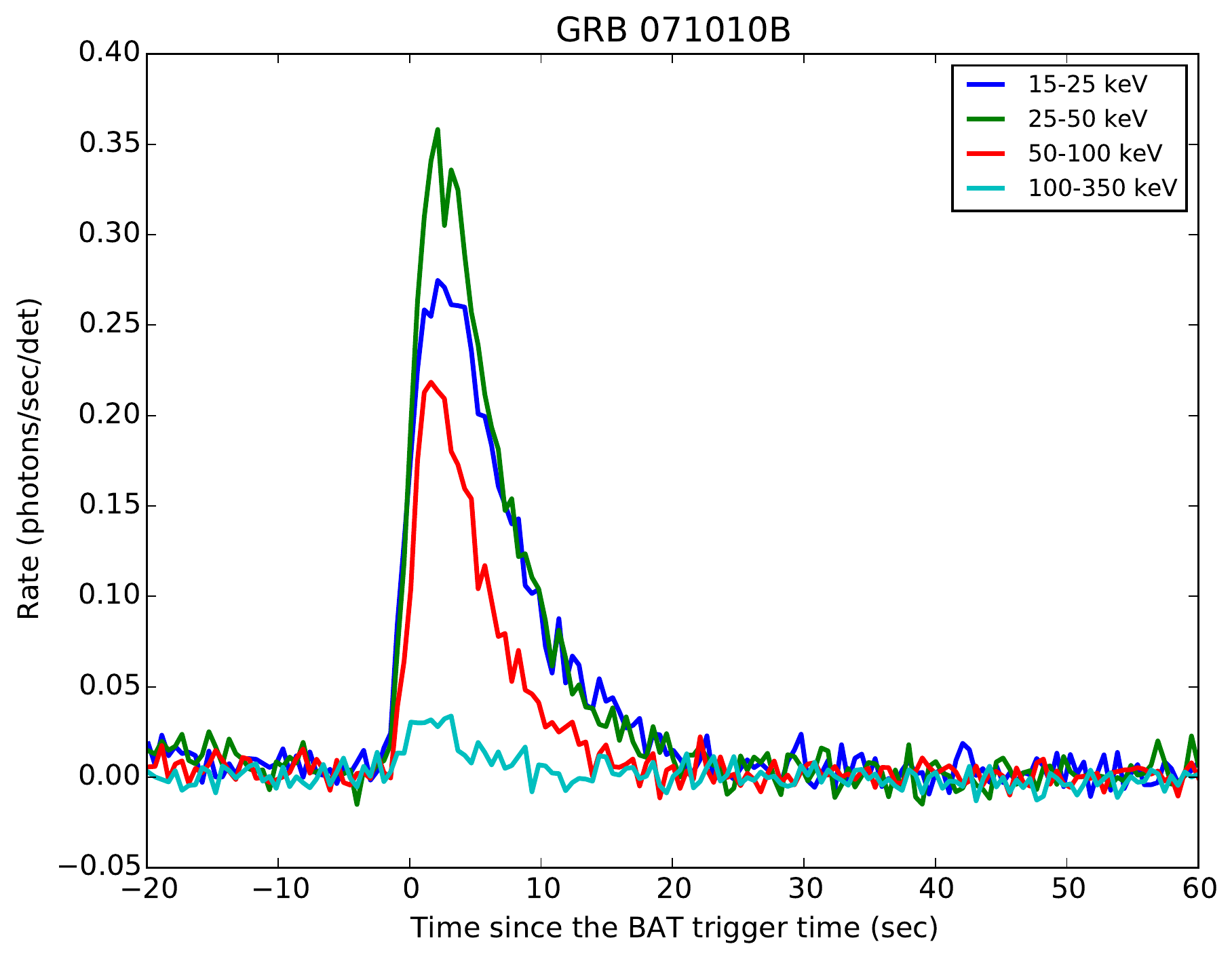}
\caption{Gamma-ray burst lightcurve of a sample burst GRB 071010B in four BAT energy bands.}
\label{grb_lightcurve}
\end{figure}

In order to catch a high-z GRB during the early bright afterglow phase,
we need to observe it as quickly as possible with a large optical observatory.
Large optical telescopes are highly over-subscribed and have limited
time to follow up GRBs. In this context, ability to screen high-z GRBs is
very important. There have been several attempts to screen high-z GRBs using
promptly available high-energy data~\cite{Grupe2007,Ukwatta2008, Ukwatta2009, Morgan2012}
The primary motivation behind these efforts is to select high-z burst early
and facilitate rapid follow-up. However, the available prompt GRB redshift estimators
are not very accurate and hence have never been adapted for wider use. In this work,
we set out to address this problem using features extracted from preprocessing based
on wavelet analysis.

\section{Methodology}\label{methodology}
\subsection{Data Selection}

The Swift mission comprises three major onboard instruments: Burst Alert Telescope (BAT),
X-ray Telescope (XRT) and UV Optical Telescope (UVOT)~\cite{Gehrels2004}.
BAT is a soft gamma-ray wide field instrument sensitive to photons in the energy
range 15 keV to 350 keV and it is the GRB discovery instrument. Once BAT discovered a GRB and
determined its sky position, the Swift satellite will slew to the location of the burst
so that the narrow field instruments, XRT and UVOT can immediately start observing the afterglow.
Fig.~\ref{grb_lightcurve} shows an example lightcurve of a GRB in four BAT
energy bands.

\begin{figure}[!t]
\centering
\includegraphics[width=3.5in]{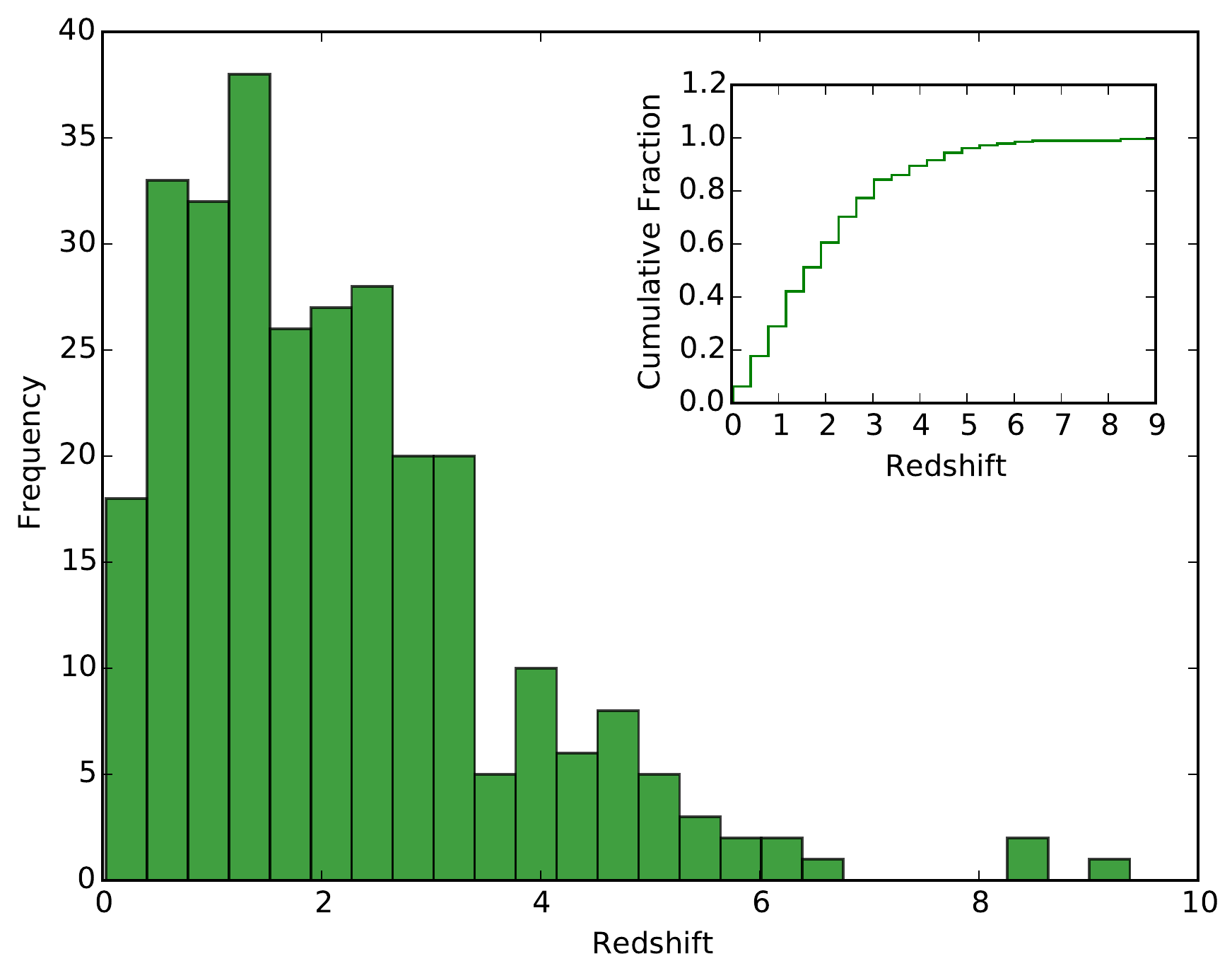}
\caption{Redshift distribution for a sample of 288 Swift GRBs.}
\label{zhisto}
\end{figure}

For our analysis, we have selected 288 long duration Swift GRBs with redshift
measurements. We have not used short duration GRBs because they are known to be
low redshift and straightforward to classify. We are interested in classifying
long duration GRBs that can have redshfits ranging from $\sim$0.03 to $\sim$9.4
in our sample. The redshfit distribution of the burst sample is shown in Fig.~\ref{zhisto}.

\subsection{GRB Image Construction}

\begin{figure*}[!t]
\centering
\includegraphics[width=6.0in]{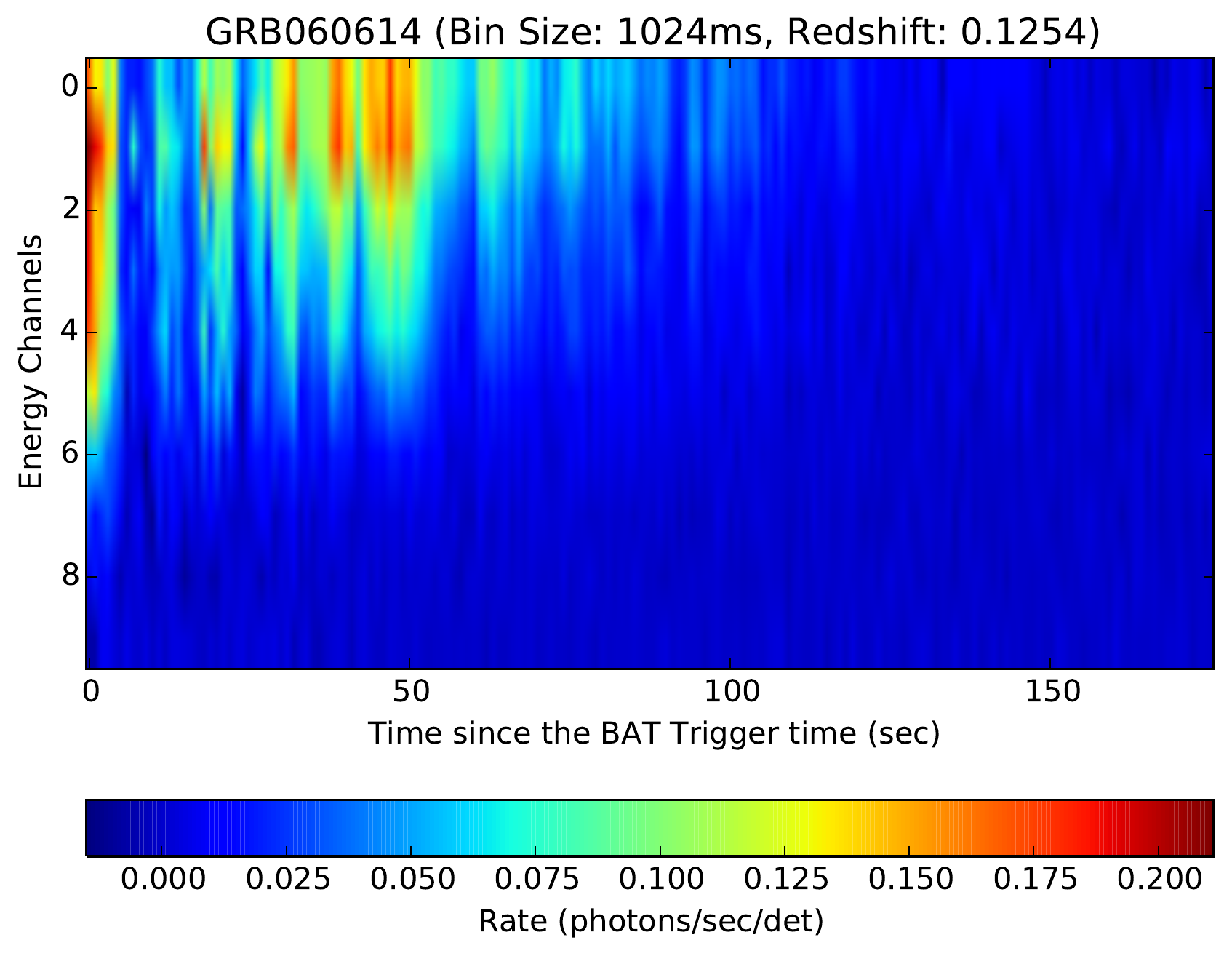}
\caption{Abstract preprocessing image for GRB 060614 constructed from 10 light curves
corresponding to energy bands in Table~\ref{ebins}.}
\label{grb_image}
\end{figure*}

As mentioned before, fitting low-parameter integrated models to light curves and spectra
may result in loss of valuable information about the time evolution and hidden features.
We address this problem by making multiple light curves in narrow
energy bands (driven by the available signal to noise ratio)
and combining them to form an abstract image. This image captures
the distribution of the photon flux in both time and energy.

In our analysis we have divided the BAT energy range into 10 logarithmic energy bins
(Table~\ref{ebins}).
An example abstract image constructed from these 10 light curves is
shown in Fig.~\ref{grb_image}. The image clearly shows both the
timing and spectral evolution of the prompt gamma-ray
emision for a single GRB. The rich structure that can be seen in this image is not
captured by the usual integrated light curves and spectra.

\begin{table}[!t]
\caption{BAT energy bands used in the analysis}
\label{ebins}
\centering
% Some packages, such as MDW tools, offer better commands for making tables
% than the plain LaTeX2e tabular which is used here.
\begin{tabular}{|c|c|}
\hline
Channel Number & Energy Band (keV)\\
\hline
0 & 15.0 -- 20.6\\
1 & 20.6 -- 28.2\\
2 & 28.2 -- 38.6\\
3 & 38.6 -- 52.9\\
4 & 52.9 -- 72.5\\
5 & 72.5 -- 99.3\\
6 & 99.3 -- 136.0\\
7 & 136.0 -- 186.4\\
8 & 186.4 -- 255.4\\
9 & 255.4 -- 350.0\\
\hline
\end{tabular}
\end{table}

\begin{figure*}[!t]
\centering
\includegraphics[width=5.0in]{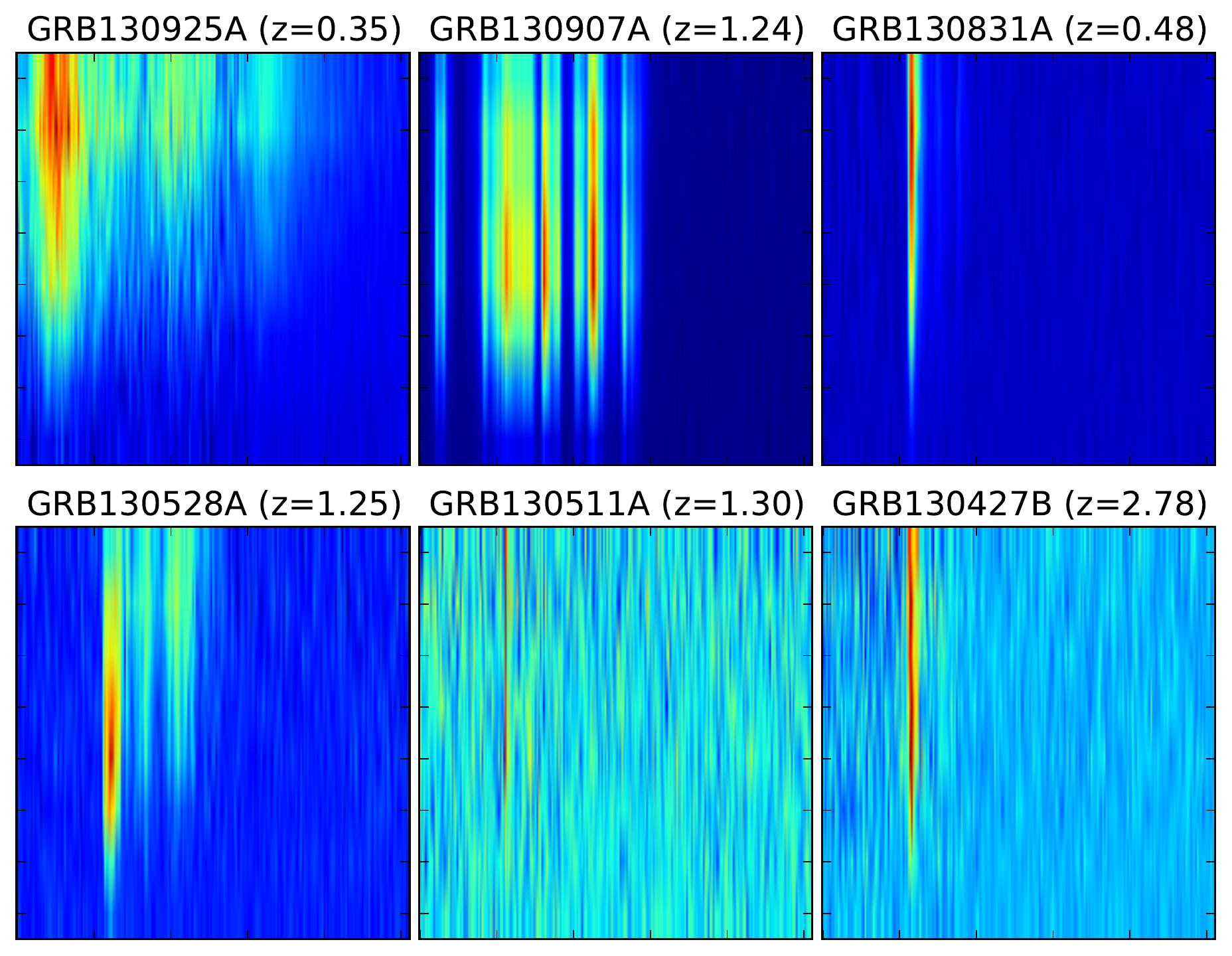}
\caption{Sample of low redshift GRBs}
\label{low_z_lc_image}
\end{figure*}

\begin{figure*}[!t]
\centering
\includegraphics[width=5.0in]{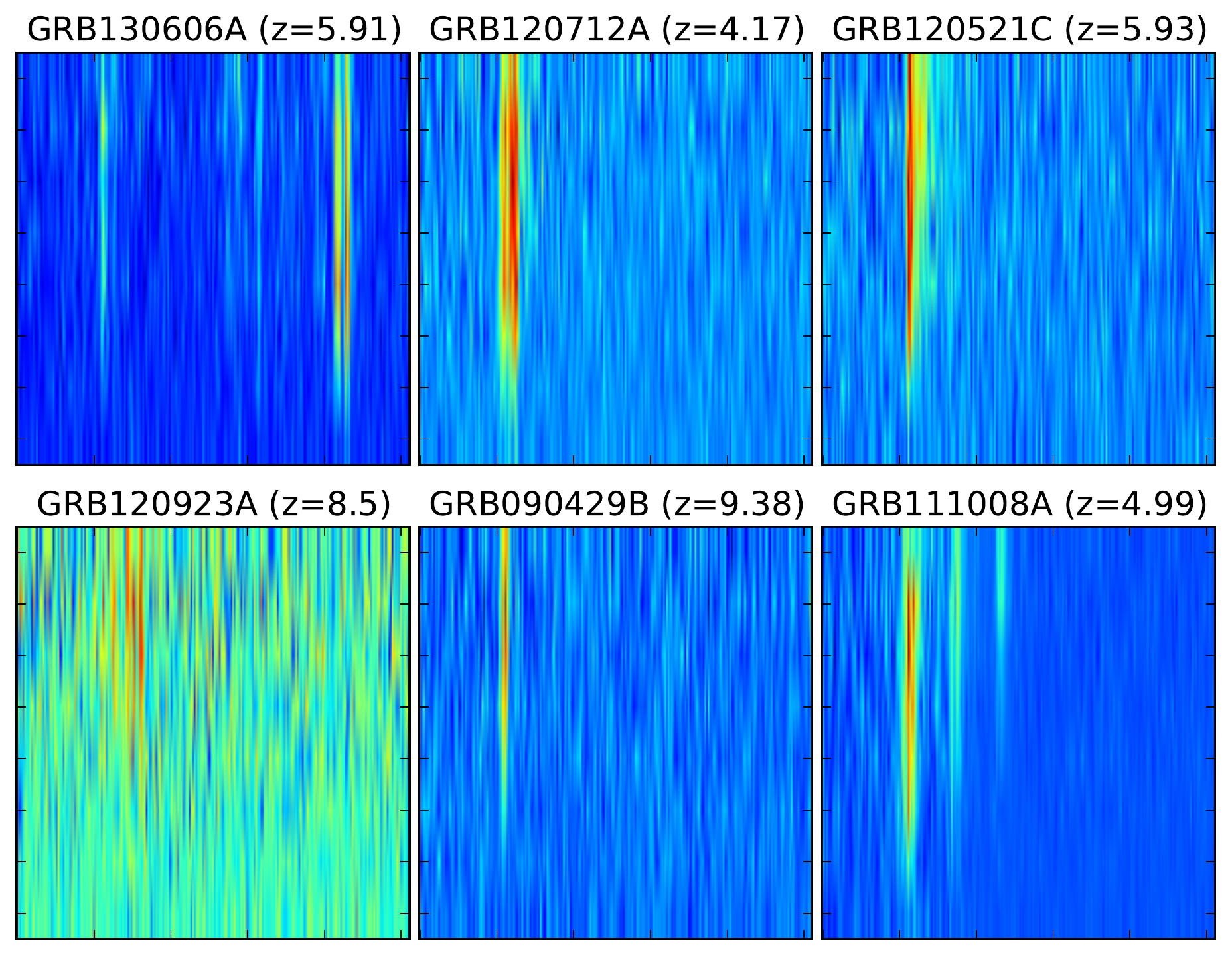}
\caption{Sample of high redshift GRBs}
\label{high_z_lc_image}
\end{figure*}

We have constructed similar images for all 288 bursts in our sample.
It is interesting to compare the images of low- and high-redshift
GRBs, and check whether there is any obvious difference between the two classes.
We consider GRBs with redshift greater than 4.0 as high redshift and
GRBs with z $<$ 4.0 as low redshift. Fig.~\ref{low_z_lc_image} depicts
six GRBs in the low redshift class and Fig.~\ref{high_z_lc_image}
shows six GRBs in the high redshift category. There are no apparent difference
that clearly stands out prior to more rigorous analysis. This is partly a result
of the intrinsic diversity of GRB light curves.
The next step is to use a machine learning algorithm to investigate whether its possible to
identify the two classes using these abstract images.

One possible approach is to input raw light curve images directly into
the classifier. This means we will feed thousands of features into
the classifier irrespective of their usefulness. This is a very
inefficient method to do classification and it is likely that providing
too many features will reduce the performance of the classifier.
We need an efficient method to extract useful features out of these
abstract images for our classification problem.

\subsection{Wavelet Analysis}

We propose to apply the wavelet transform as an efficient preprocessing method to extract useful
information from our multi-energy light curve images. Wavelet transformation allows us to analyze images
at multiple scales and extract relevant coefficients that together ``encode'' the observed structure
as a collection of time varying waves of limited duration.
Wavelets can be used to extract information about not only the fine structures in the
image but also their location. The wavelet transform is equivalent to a hierarchical
filtering process whereby the image is decomposed into progressively
finer levels of approximation and detail.
The process is repeated until the desired resolution is reached.
At each stage, the image is decomposed into four details
i.e. approximation details, horizontal details, vertical details and diagonal
details. In this analysis, we used the simplest possible wavelet which is
the Haar wavelet to analyse the abstract GRB images.

In order to compare wavelet coefficients between images we need to
treat all abstract GRB images in the same way and normalize the time dimension.
Wavelet transform requires that the image dimensions are powers of two. Otherwise
approximations will be made to fill the missing pixels. We chose to work with the
8 $\times$ 256 grid. The vertical dimension of 8 corresponds to the first 8 energy channels
given in Table~\ref{ebins}. Energy channels 8 and 9 were not used because in our sample
there is virtually no significant emission beyond 150 keV due to a sharp drop in the effective
area of the BAT detector.
The horizontal dimension of 256 corresponds to the time axis of the light curves.
We choose 56 bins before the BAT trigger time and 200 bins after the trigger time
to accommodate the time interval of significant emission for all bursts in the sample.
All light curves were binned with bin size of 1024 msec resulting in images that consist
of 2048 pixels each. The abstract GRB images shown in Fig.~\ref{low_z_lc_image}
and Fig.~\ref{high_z_lc_image} follow this prescription: 8 energy channels, 256 time intervals,
and 2048 pixels total. In this work we used python implementation of the 2D discrete wavelet
transform in the PyWavelets library (\texttt{http://www.pybytes.com/pywavelets}).

\subsection{High level classification with random forest}

The low-level abstract features from wavelet analysis are then used as input
for a machine learning algorithm that allows us to classify GRBs and select
high-redshift candidates. For this purpose we used the random forest (RF) algorithm
~\cite{Breiman2001} that has been shown to provide superior performance on many
problems in observational astrophysics. The RF algorithm creates a large number
of randomized decision trees that are then used as a voting ensemble to provide
high quality predictions.
For each tree $T$, a bootstrap sub-sample is chosen from the original training sample.
Stochasticity is injected into the process of growing trees by selecting a random subset
of $m$ features for each binary split. For a given test case, we take a majority vote
to perform classification or take an average of predictions over the ensemble to perform
regression. To reliably select one class against the others, as is the case in high-z GRB selection,
we can require a super-majority threshold that also serves to tune the location of the
final classifier on the Receiver Operating Characteristic (ROC) curve, which will be
described in the next section. For the present work we used the
python RF implementation available in the scikit-learn package
(\texttt{http://scikit-learn.org}).

\section{Results}\label{results}

For all GRBs in our sample we computed the wavelet transform up to
8 levels and extracted the corresponding coefficients.
The coefficients at each level were used for training the classifier
and evaluating its performance.
The ROC curve is a convenient way to track performance and compare classifiers and
feature sets. We calculate this curve by performing 10-fold cross-validation runs,
repeating the process 100 times, and averaging the results. The curve traces
various probability thresholds for the classification. The perfect classifier
has zero false positive rate and 100\% true positive rate, which corresponds
to the upper left corner of the diagram. Fast rising ROC curve is generally better
than a slow rising one. The area under the curve gives us a rough
measure of classification performance. An ideal classifier has the area
of 1, while random selection without a classifier yield on average an area of 0.5.

\begin{figure}[!t]
\centering
\includegraphics[width=3.5in]{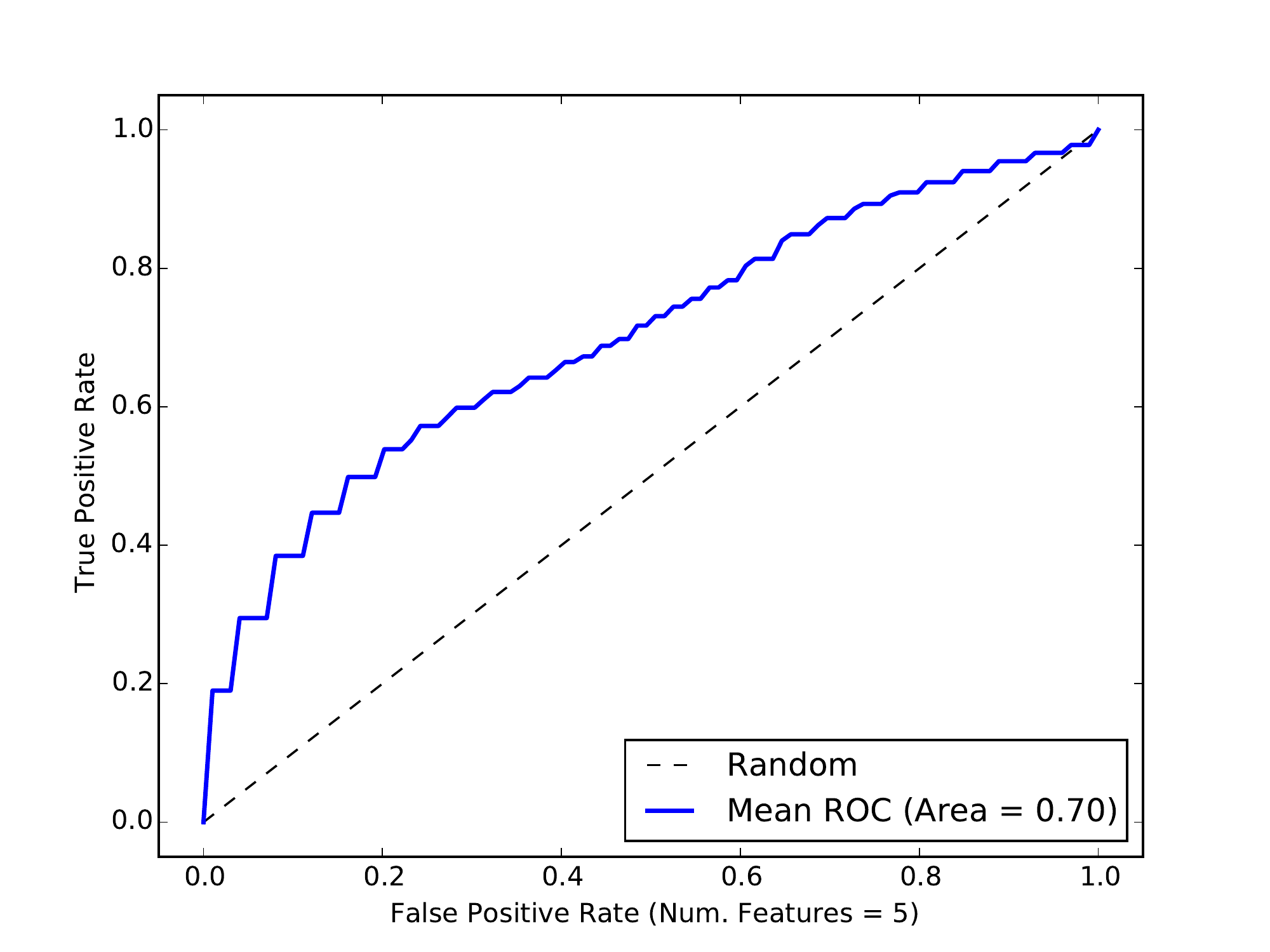}
\caption{ROC Curves for features derived from fitting BAT light curves and spectra.}
\label{bat_roc}
\end{figure}

RF classifiers take a number of parameters that can be tuned to improve
performance. The most important parameters are the number
of trees in the forest, leaf node size, and the size of the subset of
features that will be used for node splitting ($m$). A simple parameter
search was performed to establish that 500 trees with at least 15 training
samples per node, and $m$=4 random features per split gives
the best performance on the present problem.

%\begin{figure}[!t]
%\centering
%\includegraphics[width=3.5in]{GRB060614_upto_level_rec_lc_image.pdf}
%\caption{Reconstructed images from wavelet coefficients up to various level.}
%\label{upto_level_rec}
%\end{figure}

%\begin{figure}[!t]
%\centering
%\includegraphics[width=3.5in]{GRB060614_level_rec_lc_image.pdf}
%\caption{Reconstructed images using only wavelet coefficients from a level.}
%\label{only_level_rec}
%\end{figure}

\begin{figure*}[!t]
\centering
\includegraphics[width=5.0in]{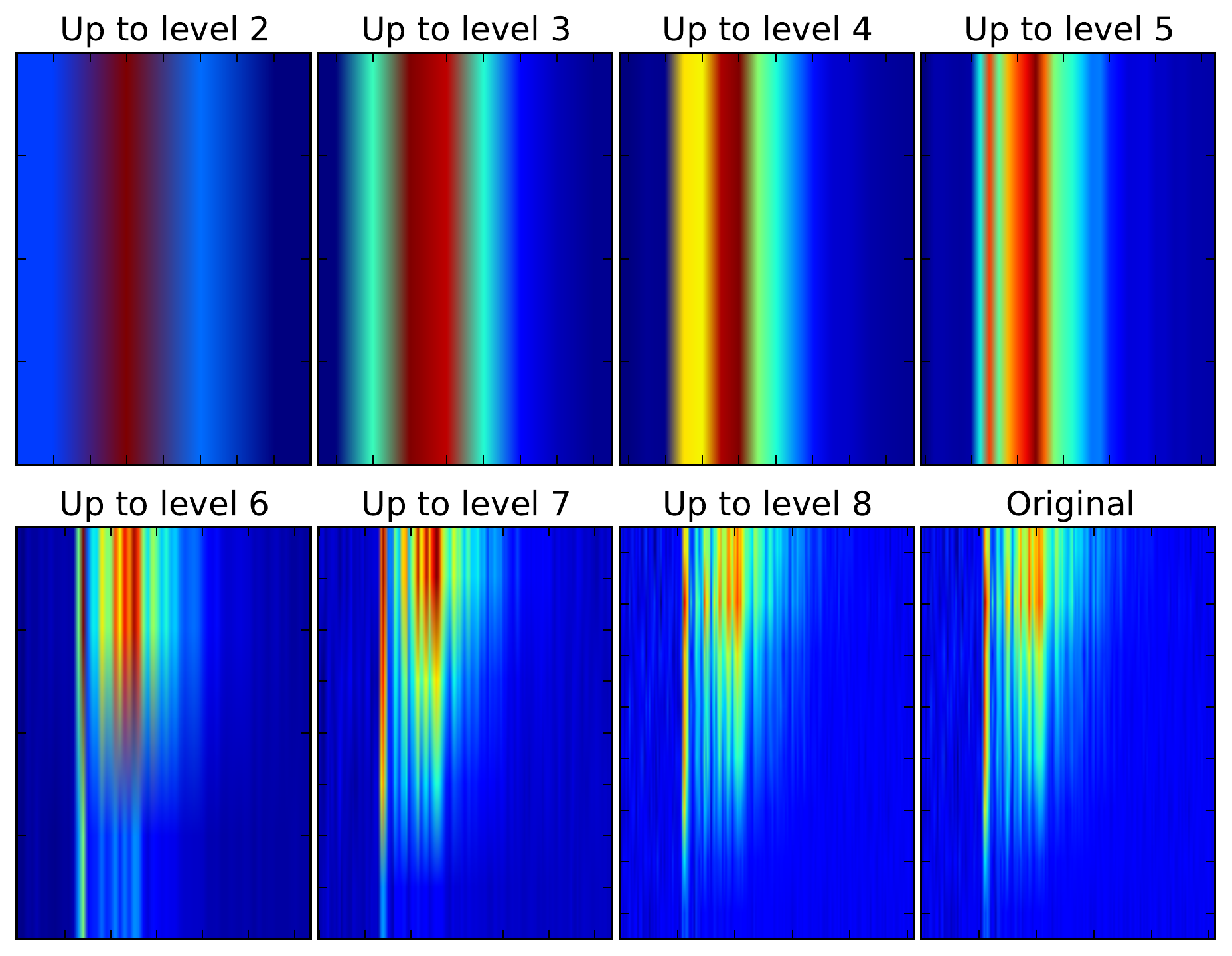}
\caption{Light curve images reconstructed using all wavelet coefficients up to a specified level.}
\label{upto_level_rec}
\end{figure*}

\begin{figure*}[!t]
\centering
\includegraphics[width=5.0in]{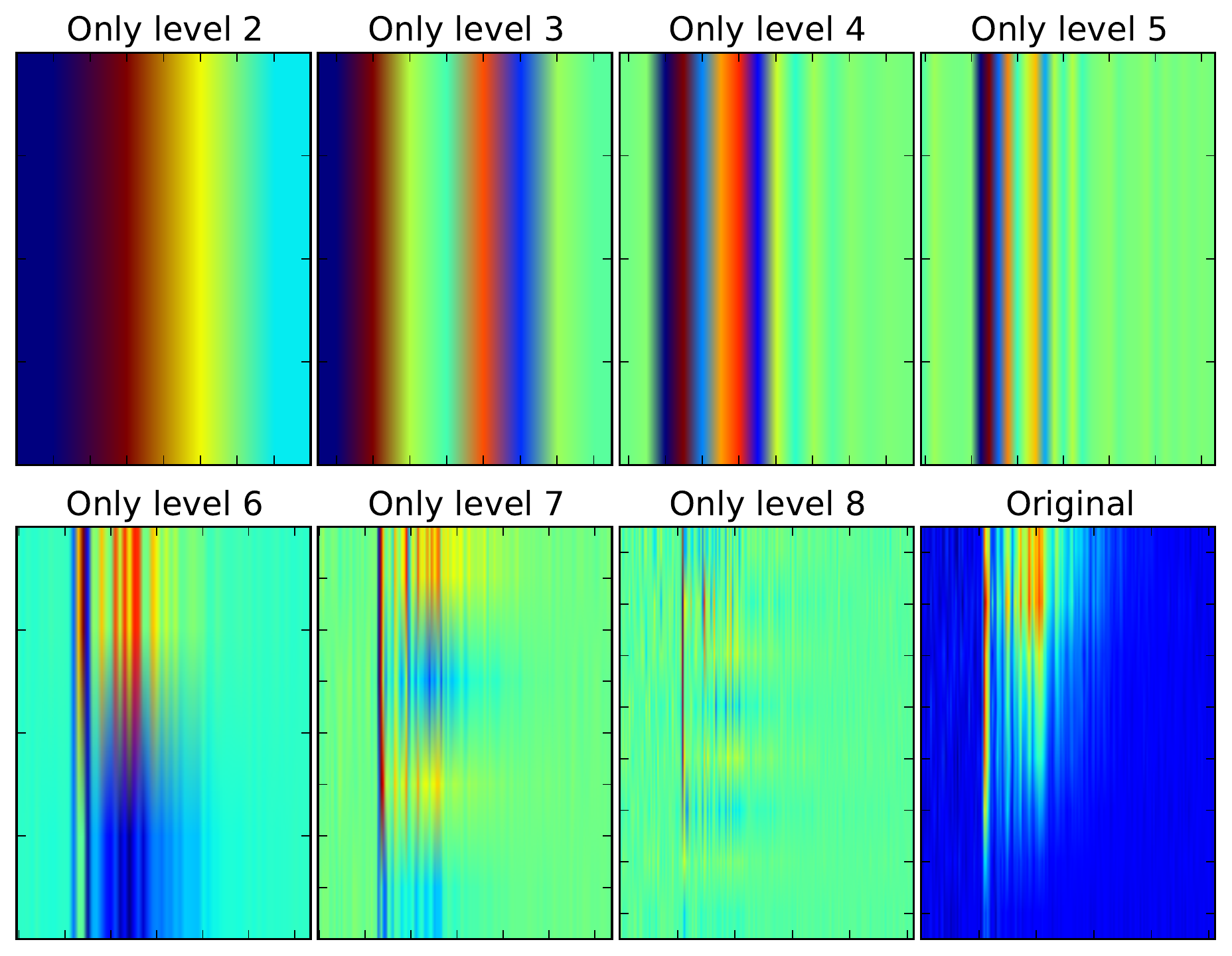}
\caption{Light curve images reconstructed using only a single level of wavelet coefficients.}
\label{only_level_rec}
\end{figure*}

\begin{figure*}[!t]
\centering
\includegraphics[width=5.5in]{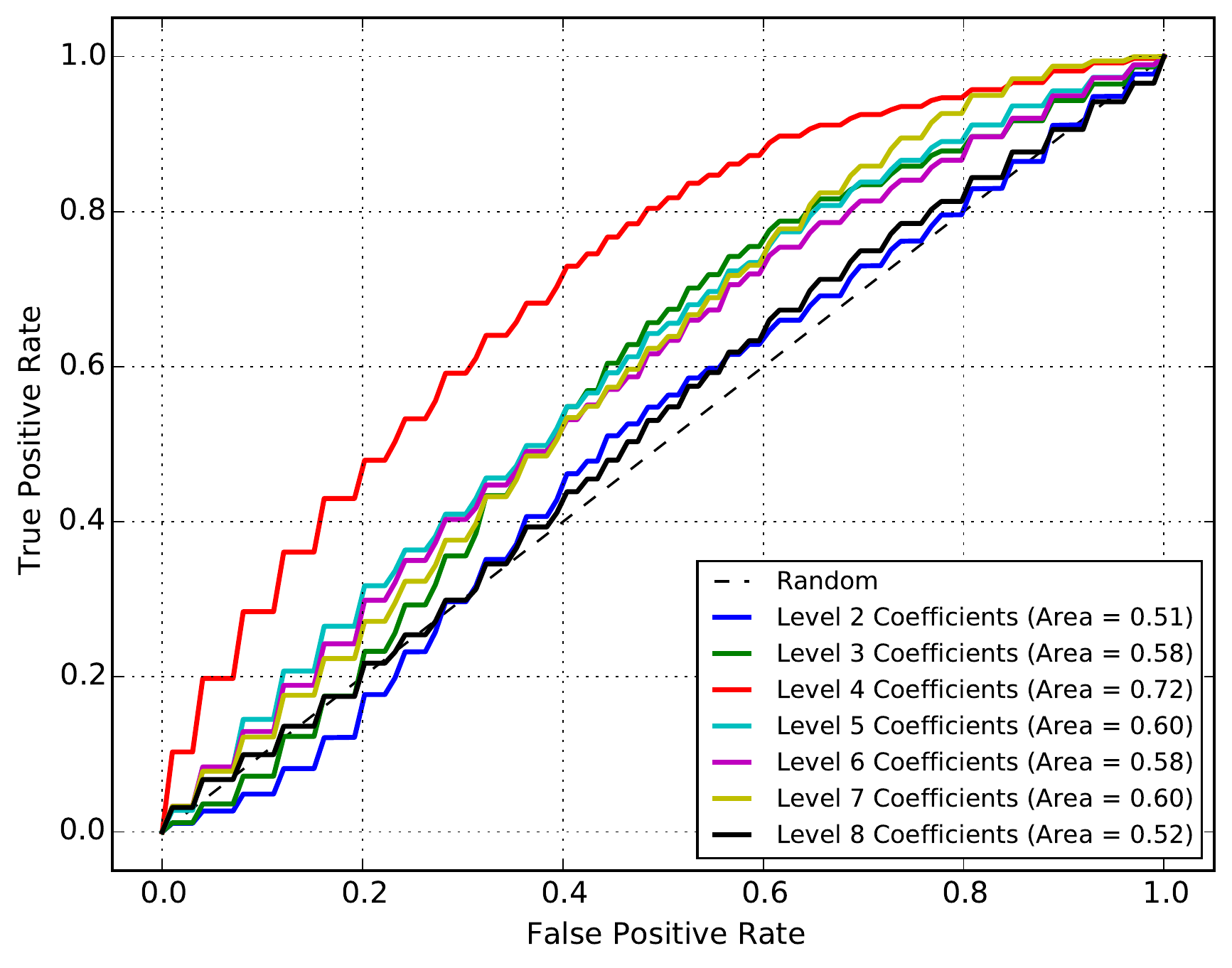}
\caption{ROC curves corresponding to detail coefficients of various levels of the wavelet
transform. Level 4 coefficients carry most information about
the redshift information encoded in the abstract GRB images.}
\label{multi_roc_curve_levels}
\end{figure*}

As a reference, we first computed ROC curves using several standard GRB
measurements obtained using the BAT data:

\begin{enumerate}
  \item T90, which is the time interval that contains 90\% of the burst
fluence centered on the mid point (i.e. starting at 5\%)
  \item Fluence which is the time integrated flux over the burst duration
  \item 1-second peak flux
  \item Spectral Index from fitting either a power-law (PL) function or
cut-off power-law (CPL) function
  \item Numerical value 0 or 1 depending on the spectral index comes from
PL fit or CPL fit
\end{enumerate}

The resulting ROC curve for these features is shown in Fig.~\ref{bat_roc}.
The curve displays a relatively fast rise early on and then flattening
around the middle. The area under the curve is 0.70.

The next step is to apply the wavelet transform and extract coefficients for all
abstract GRB images in the sample. In order to preserve features at different scales
we applied the transform up to 8 levels, which results in relatively large number of
coefficients. Figure~\ref{upto_level_rec} shows the wavelet reconstruction of the abstract
image for GRB 060614 using wavelet coefficients up to a given level. It is clear that
when we increase the level, the level of detail increases and at level 8 we recover
the original image. In order to reduce the number of coefficients and still preserve
the most important features, at each level we selected only the detail coefficients.
To illustrate various features probled at each level, we plotted the
reconstructed image of GRB 060614 using only coefficients at a given level.
This is shown in Fig.~\ref{only_level_rec}.

We trained our classifier using horizontal details, vertical details and diagonal
details at each level of the wavelet transform. The ROC curves
corresponding to various levels are shown in Fig.~\ref{multi_roc_curve_levels}.
The number of wavelet coefficients at each level is listed in Table~\ref{numcoeff}.
According to Fig.~\ref{multi_roc_curve_levels}, the detail coefficients
at level 4 carry most information on the redshift of the burst.

\begin{table}[!t]
\caption{Number of wavelet coefficients at each level of detail.}
\label{numcoeff}
\centering
\begin{tabular}{|c|c|c|}
\hline
Level & Number of Coefficients & Area\\
\hline
2 & 6 & 0.51\\
3 & 12 & 0.58\\
4 & 24 & 0.72\\
5 & 48 & 0.60\\
6 & 96 & 0.58\\
7 & 384 & 0.60\\
8 & 1536 & 0.52\\
\hline
\end{tabular}
\end{table}

\begin{figure*}[!t]
\centering
\includegraphics[width=5.5in]{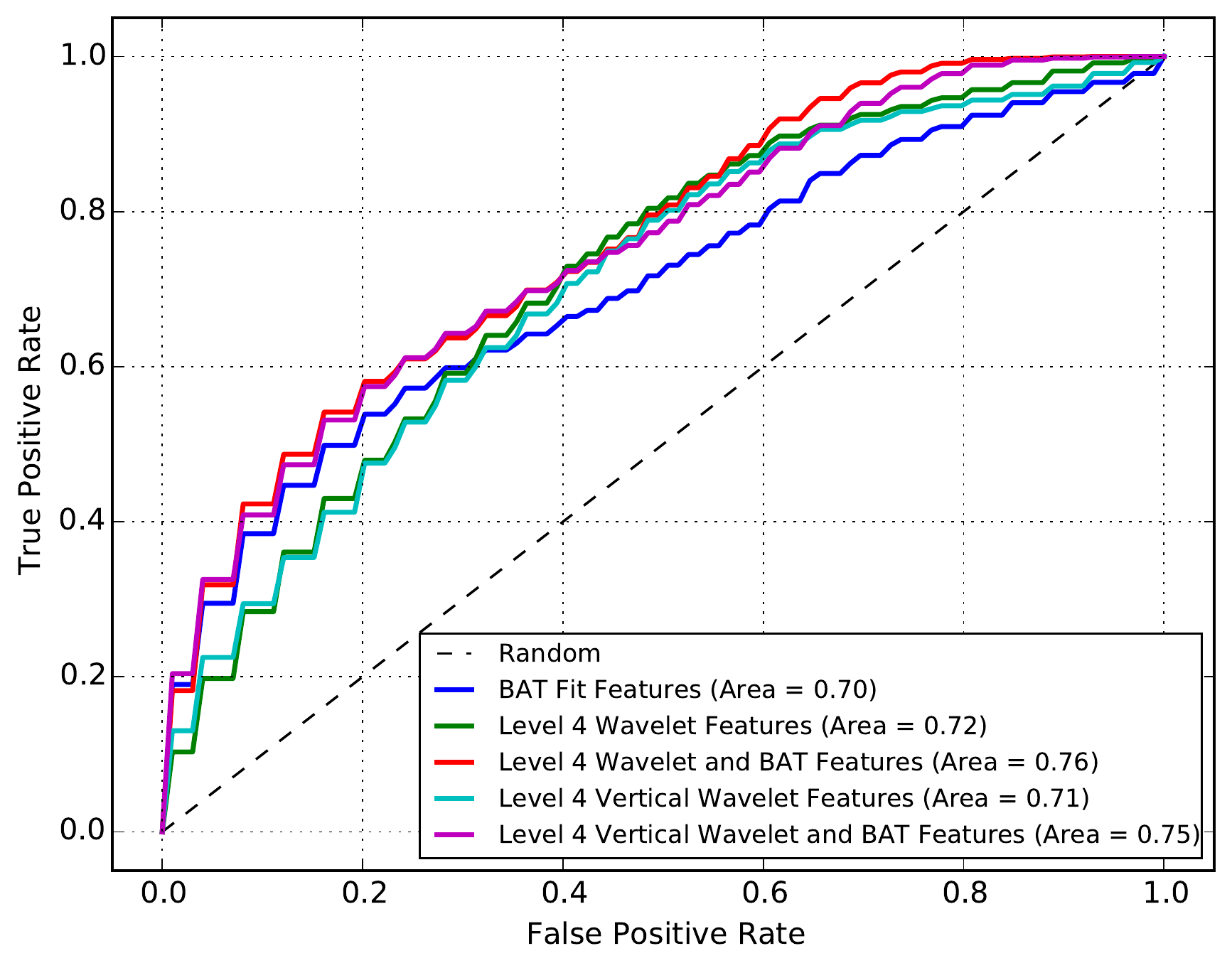}
\caption{Comparison of ROC curves for various feature selection.}
\label{multi_roc_curve}
\end{figure*}

It is important to note that the abstract
GRB images we used to select high-redshift bursts do not carry direct information
about either the duration or the energy distribution of the event.
The wavelet transform captures the relative strength of various structures in the
images and their location relative to the BAT trigger time. With level 4 detail
coefficients the ROC area is 0.72. This is slightly better than the value obtained with
features derived from fitting the BAT light curves and spectra. We therefore select
24 wavelet coefficients at level 4 as our features for the classifier.

However, a close inspection of level 4 detailed coefficients revealed that
almost exclusively the horizontal detail and diagonal detail coefficients are
equal to zero. This implied that almost all relevant information is included
in the 8 vertical detail coefficients. This is evident in Fig.~\ref{multi_roc_curve}.
Here we show a comparison of the ROC curves for all 24 level 4 coefficients and
8 level 4 vertical detail coefficients. The area is 0.72 for the former compared to 0.71
for the latter. Despite a slight performance hit, the shape of the two curves
is almost the same.

It is also interesting to see whether we can improve the performance
of the classifier by combining features from both wavelet analysis and model fitting.
The corresponding ROC curves are also shown in Fig.~\ref{multi_roc_curve}.
By combining the conventional BAT features with 8 level 4 vertical detail coefficients
we can increase the area under the ROC curve to 0.75. If we add the remaining level 4
coefficients, the area increases to 0.76. It is clear that wavelet coefficients
do capture some information about the redshift that is not included
in features derived from fitting BAT light curves and spectra.

\section{Discussion}\label{discussion}

\begin{figure*}[!t]
\centering
\includegraphics[width=5.2in]{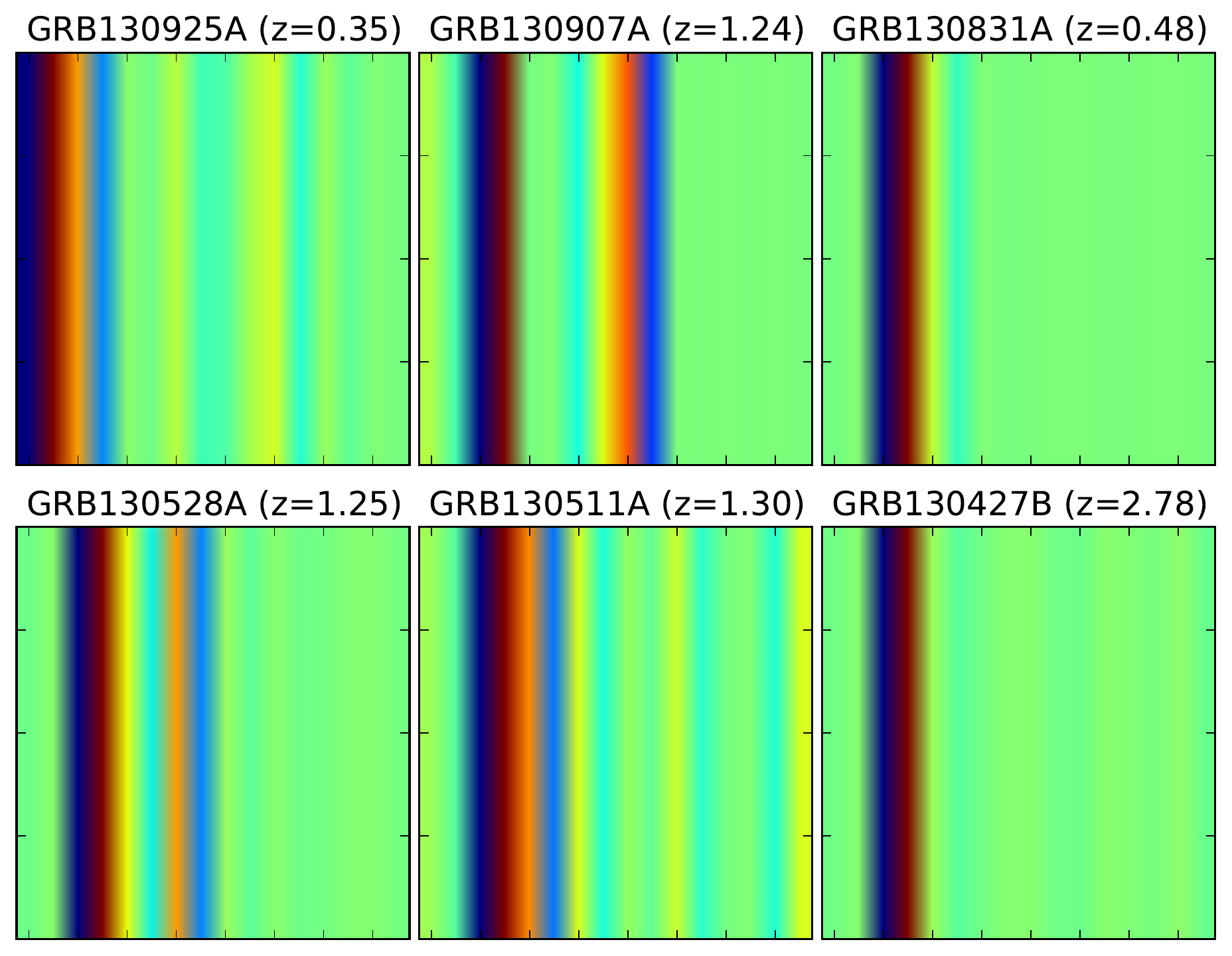}
\caption{Sample of low-redshift GRB images reconstructed from level
4 wavelet coefficients.}
\label{low_z_rec_lc_image}
\end{figure*}

\begin{figure*}[!t]
\centering
\includegraphics[width=5.2in]{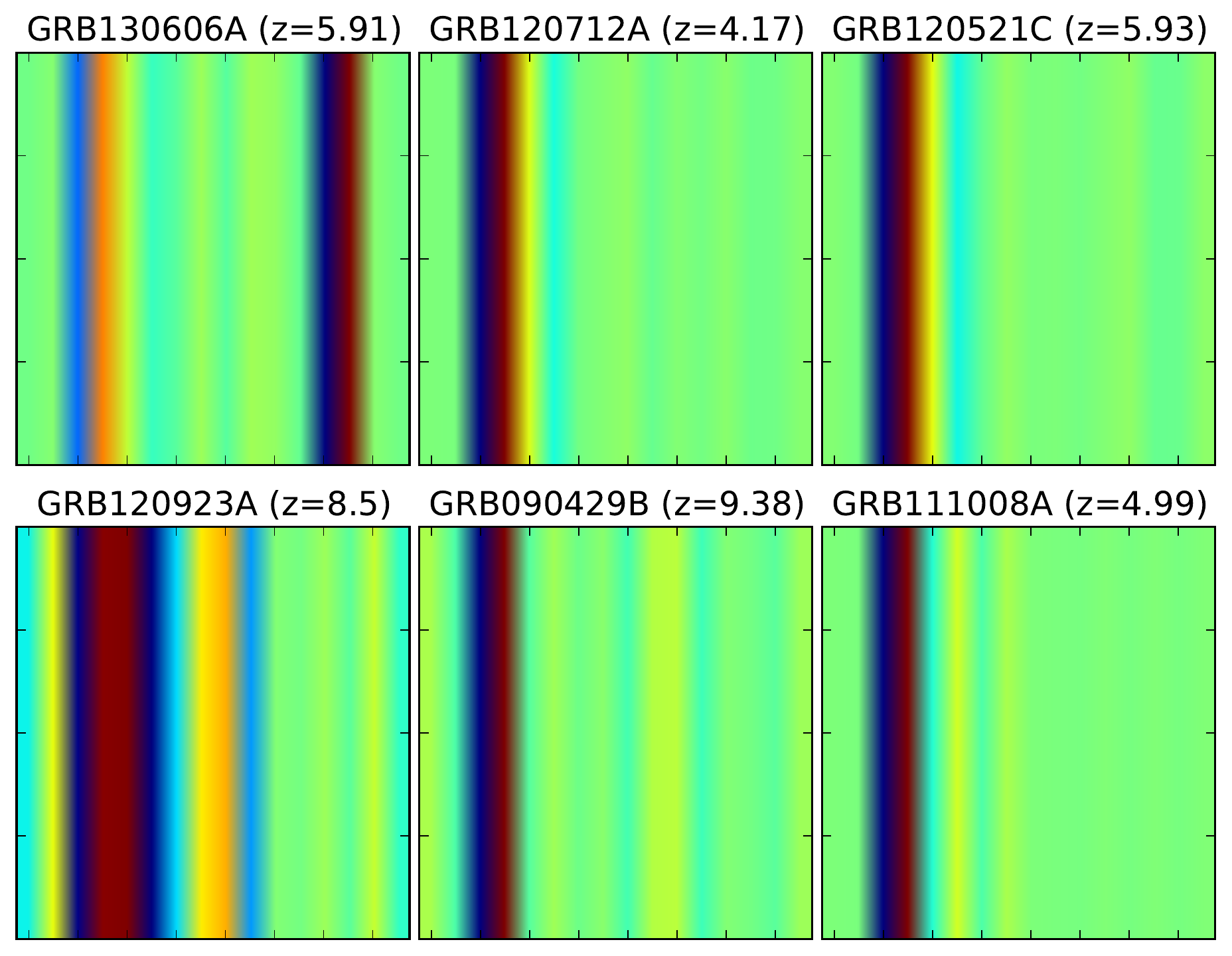}
\caption{Sample of high-redshift GRB images reconstructed from level
4 wavelet coefficients.}
\label{high_z_rec_lc_image}
\end{figure*}

It is an open question how much information about GRB distances (redshifts)
can be recovered from prompt gamma-ray emission.
We expect to see the effects of the cosmological time dilation and
energy shift signatures. This implies the existence
of some GRB property which is constant from burst to burst.
Identification of such a constant feature make GRBs potential standard candles
and that has profound implication
for GRB studies as well as cosmological studies.
From Fig.~\ref{multi_roc_curve_levels} and Fig.~\ref{multi_roc_curve} it is
clear that there exist some signal in the GRB prompt emission
that specify its redshift.

In our analysis, we have identified 8 level 4 vertical detail coefficients that
contain most of the information about the GRB redshift. This is
a dimensionality reduction of more than two orders of magnitude from 2048 to 8 numbers.
While from Fig.~\ref{multi_roc_curve_levels} it is evident that some
detail coefficients at other levels also carry information on redshift,
feeding all these coefficients blindly reduces the performance of the
classifier. On the other hand, it is not straightforward to
identify important coefficients between levels.
This is a topic for future studies.

Our analysis demonstrates the existence of the ``redshift signal''
in the GRB prompt gamma-ray emission. However, it does not reveal
which physical attributes carry this signal. In order to shed some
light on this issue we plotted the reconstructed abstract GRB images
using only level 4 coefficients for the two redshift classes shown in
Fig.~\ref{low_z_rec_lc_image} and Fig.~\ref{high_z_rec_lc_image}.
The GRBs shown here are the same as in
Figures~\ref{low_z_lc_image} and \ref{high_z_lc_image}.
These figures suggest that level 4 coefficients register
structures along the time axis much more than along the energy axis.
However, we cannot dismiss the energy structures
as unimportant because coefficients at other levels also
contain the redshift signal.

The high-z classification presented in this paper is a case study
to illustrate the utility of the analysis method. For GRBs
this method can be used to identify other classes of GRBs such
as dark bursts (burst without afterglows), magnetically
dominated GRBs, burst's in different interstellar
environments, and more. It can also be used to investigate
connections between the prompt emission and afterglows of GRBs.
In fact, any observation of astrophysical phenomenon that provides
light curves and spectra can be studied using this method.

%\begin{figure*}[!t]
%\centering
%\includegraphics[width=5.0in]{low_z_rec_lc_image.pdf}
%\caption{Sample of low redshift GRBs}
%\label{low_z_rec_lc_image}
%\end{figure*}

%\begin{figure*}[!t]
%\centering
%\includegraphics[width=5.0in]{high_z_rec_lc_image.pdf}
%\caption{Sample of high redshift GRBs}
%\label{high_z_rec_lc_image}
%\end{figure*}

\section{Conclusion} \label{conclusions}

We have introduced a novel method to compare time-resolved astronomical
spectra. The method involves creation of an abstract image
that captures both time and energy evolution of the photon stream
from astronomical sources. The wavelet transform is used
to extract important features relevant for the
machine learned classification or regression problem at hand.

As a case study, we have applied this method to Swift
GRB data and investigated whether the GRB prompt
gamma-ray emission can provide information about burst redshift.
We found that indeed there is a significant amount of information
about the redshift in the GRB prompt gamma-ray emission.
In addition, we show that by constructing an abstract image
and computing wavelet coefficients,
we can gain information that was not present in
features extracted from model fits.


\section*{Acknowledgment}
This work was funded by the US Department of Energy. We acknowledge
support from the Laboratory Directed Research and Development program at
Los Alamos National Laboratory.



%\bibliographystyle{IEEEtran}

\end{document}